\documentclass[twocolumn,twoside,slac_two]{revtex4}
\usepackage{graphicx}
\usepackage{fancyhdr}
\begin{document}

\title{The First 100 LAT Gamma-Ray Bursts: A New Detection Algorithm and Pass 8}

%

\author{G.Vianello, N.Omodei}
\affiliation{Stanford University, Stanford, CA, United States}
\author{on behalf of the Fermi/LAT collaboration}




\begin{abstract}
Observations of Gamma-Ray Bursts with the {\it Fermi} Large Area Telescope have prompted theoretical advances and posed big challenges in the understanding of such extreme sources, despite the fact that GRB emission above 100 MeV is a fairly rare event. The first {\it Fermi}/LAT GRB catalog, published a year ago, presented 28 detections out of ~300 bursts detected by the Fermi Gamma-Ray Burst Monitor (GBM) within the LAT field of view. Building on the results from that work and on recent development in the understanding of the systematic errors on GBM localizations, we developed a new detection algorithm which increased the number of detections by 40\%. Even more recently the development of the new event analysis for the LAT ("Pass 8") has increased the number of detections within the first 3 years of the mission to 45, up 50\% with respect to the published catalog. The second LAT GRB catalog, in preparation, will cover more than 6 years of the mission and will break the barrier of 100 detected GRBs, a more than 20-fold improvement with respect to observations before the {\it Fermi} era in the same energy range. We will review the main features of the new algorithm, as well as preliminary results from this investigation.

\end{abstract}

\maketitle

\thispagestyle{fancy}


\section{High-energy emission from Gamma-Ray Bursts}
\label{intro}
{\it Fermi}/LAT observations are uncovering new and unexpected properties of the high-energy emission from Gamma-Ray Bursts (GRBs) shedding light on physics mechanisms, such as particle acceleration and emission processes, in ultra-relativistic regime. The first LAT GRB catalog~\citep{2013ApJS..209...11A} contained 35 bursts, with 28 detected above 100 MeV with the standard likelihood analysis and 7 with the LAT Low-Energy technique (LLE). It established new high-energy features of GRBs, namely: 
\begin{itemize}
\item Additional power-law component during prompt emission: the prompt emission of most GRBs have been successfully described in the past with the Band function~\citep{Band:93}. In many bright GRBs observed by the LAT an additional power-law component is required to account for high-energy data.
\item Delayed onset: the emission above 100 MeV is systematically delayed with respect to the low-energy emission seen in the keV--MeV energy range.
\item Extended duration: the emission above 100 MeV is also systematically longer than the prompt emission, and decays smoothly as a power law with typical decay index of $-1$, pointing to a different physical origin with respect to the spiky prompt emission.
\end{itemize}
Early afterglow models \citep{2009arXiv0905.2417K, 2009arXiv0910.2459G} could explain the observed time decay, the delayed onset as the outflow deceleration time scale, and the lack of variability. Hadronic models could explain these features as well: the onset delay could be the time to produce electromagnetic cascades \citep[e.g.,][]{2006NJPh....8..122D, Gupta:07, 2007ApJ...671..645A} or the time required to accelerate, accumulate, and cool down relativistic protons via proton-synchrotron emission in a very strong magnetic field \citep{2009arXiv0908.0513R}. Furthermore, energy-dependent delays would be expected in proton-synchrotron models where the cooling break shifts to lower energies at later times. On the other hand, we have analyzed the record-breaking GRB 130427A finding that, while several features of the ``prompt'' emission are in agreement with some internal-shocks models, other key details such as the estimated Lorentz factors for the colliding shells contradict expectations, possibly calling for different scenarios~\citep{Preece03012014}. Also, the maximum energy of the photons in conjunction with a featureless high-energy light curve rules out both synchrotron and Synchrotron Self Compton from Fermi-accelerated particles as emission mechanism, a radical departure from the standard external-shock model for the LAT emission~\citep{Ackermann03012014}. New observations and insights are hence needed to foster further theoretical developments and overcome these difficulties. 

Our new specialized analysis described in Sect.~\ref{sec:ltft} allows the  detection of faint high-energy GRB counterparts, increasing the efficiency of detection by more than 50\% and, when used in conjuction with the new ``Pass 8'' event selection, yielding more than 100 bursts over the time span of the {\it Fermi} mission. The analysis and characterization of this new sample is in progress. When completed, it will provide the needed new insights. In particular, we will be able to settle some open questions which we could not firmly establish in the first catalog due to the limited statistics: 
\begin{itemize}
\item The existence of a separate population of hyper-energetic events, characterized by a ratio between high- and low-energy fluence much larger than the others.
\item The high-energy emission for all but 2 GRBs in the sample decays as  $t^{-1}$ at late times, which in the context of the fireball model favors an adiabatic expansion. The other two GRBs instead decay as expected from a radiative regime ($t^{-1.5}$) \citep{1976PhFl...19.1130B,1997ApJ...490..772K}, but they suffered from less-than-optimal observing conditions. Is there a class of truly radiative fireballs? 
\item One of the GRBs in the sample presented a high-energy cutoff in the extra power-law component, while another one in the low-energy Band component. How common are spectral cutoffs? Do bursts with a cutoff have any other peculiar feature? 
\end{itemize}

Analyzing the new enlarged sample will answer these questions, and will uncover new features.

\section{Pass 8 and GRBs}

\begin{figure*}[tb]
\centering
\includegraphics[width=0.7\textwidth]{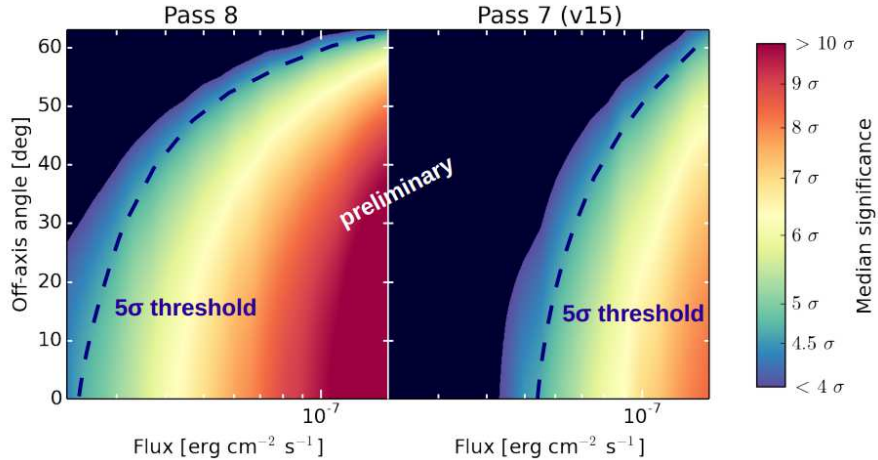}
\caption{Median significance as a function of flux and off-axis angle for Pass 8 (left) and Pass 7 (right). The dashed line is the 5$\sigma$ threshold. Pass 8 goes to much lower fluxes than Pass 7.} \label{fig:medianSignificance}
\end{figure*}

The newly-developed ``Pass 8'' analysis is an event analysis rebuilt from the ground up with respect to previous versions (``Pass 6'' and ``Pass 7''). We refer the reader to other contributions in this Symposium for an introduction and for all the details about such analysis. Here we show how it will improve the number of GRB detections. We started by performing a large number of simulations of GRBs. In particular, we fixed 10 flux values logarithmically spaced between $10^{-8}$ and $2 \times 10^{-7}$ erg cm$^{-2}$ s$^{-1}$. Since the LAT effective area is a function of the cosine of the off-axis angle $\theta$ of the GRB, we also fixed a grid of 10 values uniformly distributed in $\cos{\theta}$. For each pair (flux, $\cos{\theta}$) we performed 200 simulations. These simulations were performed with the tool \textit{gtobssim}, which accounts for the Poisson nature of the observation. For all simulations we used a fixed power-law spectrum with index $-2$, which is what is typically observed by the LAT \citep{2013ApJS..209...11A}. We then overimposed these simulated GRBs on a background taken from a real observation, and performed a likelihood analysis for each realization, recording the results. We repeated this procedure using ``Pass 7'' data at first (\textit{P7REP\_SOURCE\_V15} class), and then a preliminary version of ``Pass 8'' data (\textit{P8\_SOURCE\_V1} class). We then compared the results. In order to determine the relative sensitivity between the two datasets, we computed the median significance for each point in the (flux, $\cos{\theta}$) grid, computed as the square root of the TS value. The results are shown in Fig.~\ref{fig:medianSignificance}. It is clear that Pass 8 can detect GRBs at much lower fluxes. We stress that the actual value for a particular detection threshold (for example the $5\sigma$ one) depends on many factors (background level, duration of the GRBs, Zenith angle...) which are kept fixed in our simulations. Therefore, this is a comparative study, it is not a determination of the absolute sensitivity of Pass 8.

\section{New triggered search for Gamma-Ray Bursts}
\label{sec:ltft}

\begin{figure*}[tb]
\centering
\includegraphics[width=0.7\textwidth]{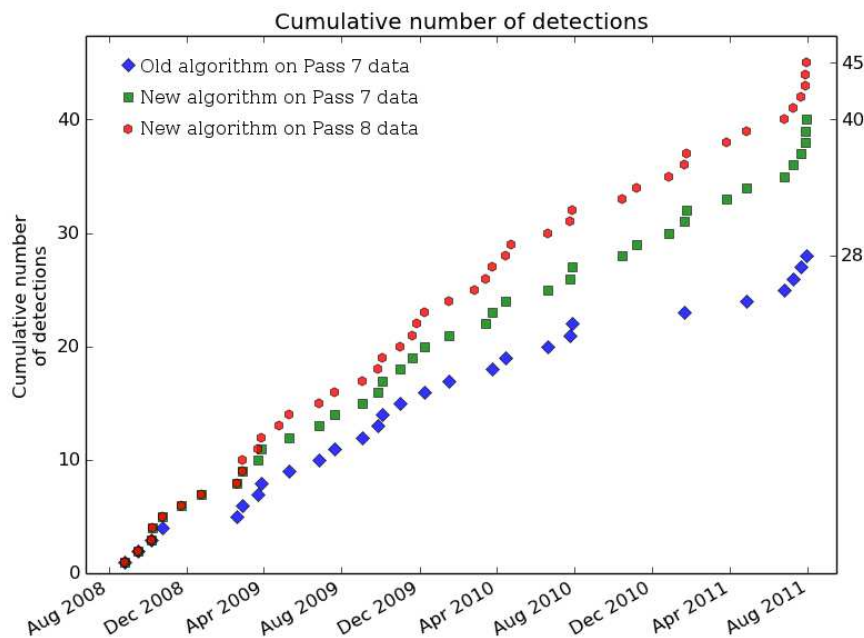}
\caption{Cumulative detections for the time span covered by the LAT GRB catalog \citep{2013ApJS..209...11A}. The new analysis yields 45\% more detections than the one used in the catalog when run on the same data (blue and green), and 60\% more with Pass 8 data.} \label{fig:ata}
\end{figure*}

Before the launch of {\it Fermi} it was estimated that the LAT would have observed 10-12 GRBs/year above 100 MeV and 6-8 above 1 GeV \citep{2009ApJ...701.1673B}. During the first 3 years, however, observations were slightly below such expectations \citep{2013ApJS..209...11A, Guetta2011}. Recently we developed a new search algorithm for GRBs, which is now up and running 24/7. As shown in Fig.~\ref{fig:ata}, it provides 45\% more detections than the algorithm used for the catalog when using Pass 7 data, and 60\% more when using the newly-developed Pass 8 data. This improvement was achieved exploiting the results of the first LAT GRB catalog, presented in Sect.~\ref{intro}. The new algorithm consists of 10  searches running in parallel over time intervals logarithmically spaced from the trigger time to 10 ks after that. For each of these time intervals, these are the steps of the new algorithm:
\begin{enumerate}
\item A trigger is received either in real-time through the GCN system\footnote{Gamma-Ray Coordinate Network, http://gcn.gsfc.nasa.gov/} or during an off-line analysis. In real-time most of the triggers are from the GBM, although also triggers from {\it Swift}, {\it INTEGRAL} and other observatories are received and processed.
\item If the trigger comes from the GBM, a ``finding map'' is produced to account for the position uncertainty. It has been recently reported that the GBM localizes GRBs with a systematic error of up to 15$^{\circ}$ \citep{2014arXiv1411.2685C}. Since this systematic error dominates over the statistical one, we always use finding maps of $30 \times 30$ deg. The finding map is essentially a TS map like the one produced by the tool \textit{gttsmap}. Indeed, a grid in equatorial coordinates and with a spacing of 0.7 deg is created covering the finding map. For each point of the grid, a likelihood analysis is performed including a new source at that position. However, points in the grid having less than 3 photons within 10 deg are not considered. This avoids running a likelihood analysis, which is computer-intensive, in regions of the finding map where there is no possibility of detecting a new point source. At the end of each likelihood a TS value is computed, and associated with the point in the grid. When all points have been processed, the maximum of the TS in the map is considered the best guess for the position of the new transient, and marked for further analysis. Note that if the trigger comes from {\it Swift} or {\it INTEGRAL}, this step is not executed, since their localization errors are much smaller than the typical size of the LAT PSF.
\item The position of the candidate transient is optimized with the tool \textit{gtfindsrc}
\item A new likelihood analysis is performed on the best position found. If the TS from this final analysis is above a certain threshold, we consider it a new detection
\item If running in real time, the results from the analysis are used by Burst Advocates to disseminate alerts to the community through GCNs.
\end{enumerate}

In all likelihood analyses involved in the sequence the likelihood model consist of the Galactic template and the Isotropic template provided by the Fermi collaboration, plus all point sources from the 2FGL source catalog. While the normalization of the two templates are left free to vary, all parameters for the 2FGL sources are kept fixed. If the algorithm finds a new candidate source with a position compatible with one of the 2FGL sources, further analysis are performed manually to distinguish between a real transients, and other phenomena such as AGN flares. The algorithm does indeed trigger on flaring sources, which are also found by other real-time algorithms such as ASP \citep{2009ApJ...701.1673B} and FAVA \citep{FAVA}.

The procedure involves a certain number of trials, which might appear large. However, all the time windows are overlapping and involve the same region on the sky, therefore these trials are not independent at all, with many photons present in many of the time windows. Thus, the effective number of trials is rather small. While the full characterization is still in progress, preliminary simulations indicates that a $5\sigma$ detection corresponds to $TS \sim 28$. 

\begin{figure*}[tb]
\centering
\includegraphics[width=0.7\textwidth]{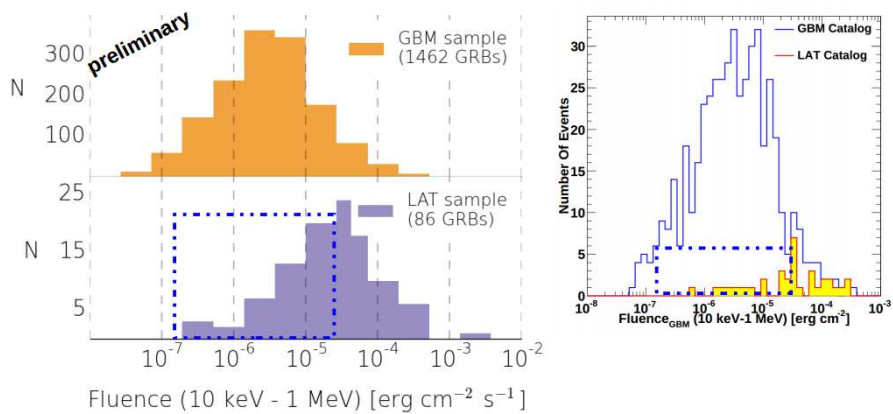}
\caption{Left panel: low-energy fluence distribution for the whole GBM sample (upper panel) and for the sample detected with the new algorithm in LAT data (lower panel). Right panel: similar plot from the first LAT GRB catalog \citep{2013ApJS..209...11A}. The blue dashed line marks the region where most of the new detections lie, showing the increased sensitivity of the new algorithm with respect to the old one.} \label{fig:fluences}
\end{figure*}

A first run of the new algorithm on $\sim 5$ years of preliminary Pass 8 data returned 86 detections. Adding the $\sim 20$ LLE-only detections, we have already reached the milestone of 100 LAT-detected GRBs. Given the current rate of detections we also expect to exceed 100 likelihood detections within the year, an impressive milestone significantly exceeding pre-launch expectations. As expected, the new detections populate the lower part of the fluence distribution, demonstrating that the new algorithm and ``Pass 8'' improve the sensitivity of the search. This is shown in Fig.~\ref{fig:fluences}.

We also note that this preliminary study is using the ``Pass 8'' Source class for all time scales. It has been shown in the past that for short time scales ($< 200 s$) using classes with larger acceptance at the expense of a larger background is beneficial. We will be using such classes in the final study, further increasing the number of detections.

\bigskip 
\begin{acknowledgments}
The \textit{Fermi} LAT Collaboration acknowledges generous ongoing support
from a number of agencies and institutes that have supported both the
development and the operation of the LAT as well as scientific data analysis.
These include the National Aeronautics and Space Administration and the
Department of Energy in the United States, the Commissariat \`a l'Energie Atomique
and the Centre National de la Recherche Scientifique / Institut National de Physique
Nucl\'eaire et de Physique des Particules in France, the Agenzia Spaziale Italiana
and the Istituto Nazionale di Fisica Nucleare in Italy, the Ministry of Education,
Culture, Sports, Science and Technology (MEXT), High Energy Accelerator Research
Organization (KEK) and Japan Aerospace Exploration Agency (JAXA) in Japan, and
the K.~A.~Wallenberg Foundation, the Swedish Research Council and the
Swedish National Space Board in Sweden.
 
Additional support for science analysis during the operations phase is gratefully acknowledged from the Istituto Nazionale di Astrofisica in Italy and the Centre National d'\'Etudes Spatiales in France.
\end{acknowledgments}

\bigskip 
\bibliography{GLAST_GRB}

\begin{thebibliography}{16}
\expandafter\ifx\csname natexlab\endcsname\relax\def\natexlab#1{#1}\fi
\expandafter\ifx\csname bibnamefont\endcsname\relax
  \def\bibnamefont#1{#1}\fi
\expandafter\ifx\csname bibfnamefont\endcsname\relax
  \def\bibfnamefont#1{#1}\fi
\expandafter\ifx\csname citenamefont\endcsname\relax
  \def\citenamefont#1{#1}\fi
\expandafter\ifx\csname url\endcsname\relax
  \def\url#1{\texttt{#1}}\fi
\expandafter\ifx\csname urlprefix\endcsname\relax\def\urlprefix{URL }\fi
\providecommand{\bibinfo}[2]{#2}
\providecommand{\eprint}[2][]{\url{#2}}

\bibitem[{\citenamefont{Ackermann et~al.}(2013)\citenamefont{Ackermann, Ajello,
  Asano, Axelsson, Baldini, Ballet, Barbiellini, Bastieri, Bechtol, Bellazzini
  et~al.}}]{2013ApJS..209...11A}
\bibinfo{author}{\bibfnamefont{M.}~\bibnamefont{Ackermann}},
  \bibinfo{author}{\bibfnamefont{M.}~\bibnamefont{Ajello}},
  \bibinfo{author}{\bibfnamefont{K.}~\bibnamefont{Asano}},
  \bibinfo{author}{\bibfnamefont{M.}~\bibnamefont{Axelsson}},
  \bibinfo{author}{\bibfnamefont{L.}~\bibnamefont{Baldini}},
  \bibinfo{author}{\bibfnamefont{J.}~\bibnamefont{Ballet}},
  \bibinfo{author}{\bibfnamefont{G.}~\bibnamefont{Barbiellini}},
  \bibinfo{author}{\bibfnamefont{D.}~\bibnamefont{Bastieri}},
  \bibinfo{author}{\bibfnamefont{K.}~\bibnamefont{Bechtol}},
  \bibinfo{author}{\bibfnamefont{R.}~\bibnamefont{Bellazzini}},
  \bibnamefont{et~al.}, \bibinfo{journal}{\apjs}
  \textbf{\bibinfo{volume}{209}}, \bibinfo{eid}{11} (\bibinfo{year}{2013}),
  \eprint{1309.4899}.

\bibitem[{\citenamefont{{Band} et~al.}(1993)\citenamefont{{Band}, {Matteson},
  {Ford}, {Schaefer}, {Palmer}, {Teegarden}, {Cline}, {Briggs}, {Paciesas},
  {Pendleton} et~al.}}]{Band:93}
\bibinfo{author}{\bibfnamefont{D.}~\bibnamefont{{Band}}},
  \bibinfo{author}{\bibfnamefont{J.}~\bibnamefont{{Matteson}}},
  \bibinfo{author}{\bibfnamefont{L.}~\bibnamefont{{Ford}}},
  \bibinfo{author}{\bibfnamefont{B.}~\bibnamefont{{Schaefer}}},
  \bibinfo{author}{\bibfnamefont{D.}~\bibnamefont{{Palmer}}},
  \bibinfo{author}{\bibfnamefont{B.}~\bibnamefont{{Teegarden}}},
  \bibinfo{author}{\bibfnamefont{T.}~\bibnamefont{{Cline}}},
  \bibinfo{author}{\bibfnamefont{M.}~\bibnamefont{{Briggs}}},
  \bibinfo{author}{\bibfnamefont{W.}~\bibnamefont{{Paciesas}}},
  \bibinfo{author}{\bibfnamefont{G.}~\bibnamefont{{Pendleton}}},
  \bibnamefont{et~al.}, \bibinfo{journal}{\apj} \textbf{\bibinfo{volume}{413}},
  \bibinfo{pages}{281} (\bibinfo{year}{1993}).

\bibitem[{\citenamefont{{Kumar} and {Barniol
  Duran}}(2009)}]{2009arXiv0905.2417K}
\bibinfo{author}{\bibfnamefont{P.}~\bibnamefont{{Kumar}}} \bibnamefont{and}
  \bibinfo{author}{\bibfnamefont{R.}~\bibnamefont{{Barniol Duran}}},
  \bibinfo{journal}{ArXiv e-prints}  (\bibinfo{year}{2009}).

\bibitem[{\citenamefont{{Ghisellini} et~al.}(2010)\citenamefont{{Ghisellini},
  {Ghirlanda}, {Nava}, and {Celotti}}}]{2009arXiv0910.2459G}
\bibinfo{author}{\bibfnamefont{G.}~\bibnamefont{{Ghisellini}}},
  \bibinfo{author}{\bibfnamefont{G.}~\bibnamefont{{Ghirlanda}}},
  \bibinfo{author}{\bibfnamefont{L.}~\bibnamefont{{Nava}}}, \bibnamefont{and}
  \bibinfo{author}{\bibfnamefont{A.}~\bibnamefont{{Celotti}}},
  \bibinfo{journal}{\mnras} \textbf{\bibinfo{volume}{403}},
  \bibinfo{pages}{926} (\bibinfo{year}{2010}), \eprint{0910.2459}.

\bibitem[{\citenamefont{{Dermer} and {Atoyan}}(2006)}]{2006NJPh....8..122D}
\bibinfo{author}{\bibfnamefont{C.~D.} \bibnamefont{{Dermer}}} \bibnamefont{and}
  \bibinfo{author}{\bibfnamefont{A.}~\bibnamefont{{Atoyan}}},
  \bibinfo{journal}{New Journal of Physics} \textbf{\bibinfo{volume}{8}},
  \bibinfo{pages}{122} (\bibinfo{year}{2006}), \eprint{arXiv:astro-ph/0606629}.

\bibitem[{\citenamefont{{Gupta} and {Zhang}}(2007)}]{Gupta:07}
\bibinfo{author}{\bibfnamefont{N.}~\bibnamefont{{Gupta}}} \bibnamefont{and}
  \bibinfo{author}{\bibfnamefont{B.}~\bibnamefont{{Zhang}}},
  \bibinfo{journal}{\mnras} \textbf{\bibinfo{volume}{380}}, \bibinfo{pages}{78}
  (\bibinfo{year}{2007}), \eprint{arXiv:0704.1329}.

\bibitem[{\citenamefont{{Asano} and {Inoue}}(2007)}]{2007ApJ...671..645A}
\bibinfo{author}{\bibfnamefont{K.}~\bibnamefont{{Asano}}} \bibnamefont{and}
  \bibinfo{author}{\bibfnamefont{S.}~\bibnamefont{{Inoue}}},
  \bibinfo{journal}{\apj} \textbf{\bibinfo{volume}{671}}, \bibinfo{pages}{645}
  (\bibinfo{year}{2007}), \eprint{0705.2910}.

\bibitem[{\citenamefont{{Razzaque} et~al.}(2009)\citenamefont{{Razzaque},
  {Dermer}, and {Finke}}}]{2009arXiv0908.0513R}
\bibinfo{author}{\bibfnamefont{S.}~\bibnamefont{{Razzaque}}},
  \bibinfo{author}{\bibfnamefont{C.~D.} \bibnamefont{{Dermer}}},
  \bibnamefont{and} \bibinfo{author}{\bibfnamefont{J.~D.}
  \bibnamefont{{Finke}}}, \bibinfo{journal}{ArXiv e-prints}
  \textbf{\bibinfo{volume}{0908.0513}} (\bibinfo{year}{2009}).

\bibitem[{\citenamefont{Preece et~al.}(2014)\citenamefont{Preece, Burgess, von
  Kienlin, Bhat, Briggs, Byrne, Chaplin, Cleveland, Collazzi, Connaughton
  et~al.}}]{Preece03012014}
\bibinfo{author}{\bibfnamefont{R.}~\bibnamefont{Preece}},
  \bibinfo{author}{\bibfnamefont{J.~M.} \bibnamefont{Burgess}},
  \bibinfo{author}{\bibfnamefont{A.}~\bibnamefont{von Kienlin}},
  \bibinfo{author}{\bibfnamefont{P.~N.} \bibnamefont{Bhat}},
  \bibinfo{author}{\bibfnamefont{M.~S.} \bibnamefont{Briggs}},
  \bibinfo{author}{\bibfnamefont{D.}~\bibnamefont{Byrne}},
  \bibinfo{author}{\bibfnamefont{V.}~\bibnamefont{Chaplin}},
  \bibinfo{author}{\bibfnamefont{W.}~\bibnamefont{Cleveland}},
  \bibinfo{author}{\bibfnamefont{A.~C.} \bibnamefont{Collazzi}},
  \bibinfo{author}{\bibfnamefont{V.}~\bibnamefont{Connaughton}},
  \bibnamefont{et~al.}, \bibinfo{journal}{Science}
  \textbf{\bibinfo{volume}{343}}, \bibinfo{pages}{51} (\bibinfo{year}{2014}),
  \eprint{http://www.sciencemag.org/content/343/6166/51.full.pdf},
  \urlprefix\url{http://www.sciencemag.org/content/343/6166/51.abstract}.

\bibitem[{\citenamefont{Ackermann et~al.}(2014)\citenamefont{Ackermann, Ajello,
  Asano, Atwood, and Axelsson}}]{Ackermann03012014}
\bibinfo{author}{\bibfnamefont{M.}~\bibnamefont{Ackermann}},
  \bibinfo{author}{\bibfnamefont{M.}~\bibnamefont{Ajello}},
  \bibinfo{author}{\bibfnamefont{K.}~\bibnamefont{Asano}},
  \bibinfo{author}{\bibfnamefont{W.~B.} \bibnamefont{Atwood}},
  \bibnamefont{and} \bibinfo{author}{\bibfnamefont{M.}~\bibnamefont{Axelsson}},
  \bibinfo{journal}{Science} \textbf{\bibinfo{volume}{343}},
  \bibinfo{pages}{42} (\bibinfo{year}{2014}),
  \eprint{http://www.sciencemag.org/content/343/6166/42.full.pdf},
  \urlprefix\url{http://www.sciencemag.org/content/343/6166/42.abstract}.

\bibitem[{\citenamefont{{Blandford} and {McKee}}(1976)}]{1976PhFl...19.1130B}
\bibinfo{author}{\bibfnamefont{R.~D.} \bibnamefont{{Blandford}}}
  \bibnamefont{and} \bibinfo{author}{\bibfnamefont{C.~F.}
  \bibnamefont{{McKee}}}, \bibinfo{journal}{Physics of Fluids}
  \textbf{\bibinfo{volume}{19}}, \bibinfo{pages}{1130} (\bibinfo{year}{1976}).

\bibitem[{\citenamefont{{Katz} and {Piran}}(1997)}]{1997ApJ...490..772K}
\bibinfo{author}{\bibfnamefont{J.~I.} \bibnamefont{{Katz}}} \bibnamefont{and}
  \bibinfo{author}{\bibfnamefont{T.}~\bibnamefont{{Piran}}},
  \bibinfo{journal}{\apj} \textbf{\bibinfo{volume}{490}}, \bibinfo{pages}{772}
  (\bibinfo{year}{1997}).

\bibitem[{\citenamefont{{Band} et~al.}(2009)\citenamefont{{Band}, {Axelsson},
  {Baldini}, {Barbiellini}, {Baring}, {Bastieri}, {Battelino}, {Bellazzini},
  {Bissaldi}, {Bogaert} et~al.}}]{2009ApJ...701.1673B}
\bibinfo{author}{\bibfnamefont{D.~L.} \bibnamefont{{Band}}},
  \bibinfo{author}{\bibfnamefont{M.}~\bibnamefont{{Axelsson}}},
  \bibinfo{author}{\bibfnamefont{L.}~\bibnamefont{{Baldini}}},
  \bibinfo{author}{\bibfnamefont{G.}~\bibnamefont{{Barbiellini}}},
  \bibinfo{author}{\bibfnamefont{M.~G.} \bibnamefont{{Baring}}},
  \bibinfo{author}{\bibfnamefont{D.}~\bibnamefont{{Bastieri}}},
  \bibinfo{author}{\bibfnamefont{M.}~\bibnamefont{{Battelino}}},
  \bibinfo{author}{\bibfnamefont{R.}~\bibnamefont{{Bellazzini}}},
  \bibinfo{author}{\bibfnamefont{E.}~\bibnamefont{{Bissaldi}}},
  \bibinfo{author}{\bibfnamefont{G.}~\bibnamefont{{Bogaert}}},
  \bibnamefont{et~al.}, \bibinfo{journal}{\apj} \textbf{\bibinfo{volume}{701}},
  \bibinfo{pages}{1673} (\bibinfo{year}{2009}), \eprint{0906.0991}.

\bibitem[{\citenamefont{{Guetta} et~al.}(2011)\citenamefont{{Guetta}, {Pian},
  and {Waxman}}}]{Guetta2011}
\bibinfo{author}{\bibfnamefont{D.}~\bibnamefont{{Guetta}}},
  \bibinfo{author}{\bibfnamefont{E.}~\bibnamefont{{Pian}}}, \bibnamefont{and}
  \bibinfo{author}{\bibfnamefont{E.}~\bibnamefont{{Waxman}}},
  \bibinfo{journal}{\aap} \textbf{\bibinfo{volume}{525}}, \bibinfo{eid}{A53}
  (\bibinfo{year}{2011}), \eprint{1003.0566}.

\bibitem[{\citenamefont{{Connaughton} et~al.}(2014)\citenamefont{{Connaughton},
  {Briggs}, {Goldstein}, {Meegan}, {Paciesas}, {Preece}, {Wilson-Hodge},
  {Gibby}, {Greiner}, {Gruber} et~al.}}]{2014arXiv1411.2685C}
\bibinfo{author}{\bibfnamefont{V.}~\bibnamefont{{Connaughton}}},
  \bibinfo{author}{\bibfnamefont{M.~S.} \bibnamefont{{Briggs}}},
  \bibinfo{author}{\bibfnamefont{A.}~\bibnamefont{{Goldstein}}},
  \bibinfo{author}{\bibfnamefont{C.~A.} \bibnamefont{{Meegan}}},
  \bibinfo{author}{\bibfnamefont{W.~S.} \bibnamefont{{Paciesas}}},
  \bibinfo{author}{\bibfnamefont{R.~D.} \bibnamefont{{Preece}}},
  \bibinfo{author}{\bibfnamefont{C.~A.} \bibnamefont{{Wilson-Hodge}}},
  \bibinfo{author}{\bibfnamefont{M.~H.} \bibnamefont{{Gibby}}},
  \bibinfo{author}{\bibfnamefont{J.}~\bibnamefont{{Greiner}}},
  \bibinfo{author}{\bibfnamefont{D.}~\bibnamefont{{Gruber}}},
  \bibnamefont{et~al.}, \bibinfo{journal}{ArXiv e-prints}
  (\bibinfo{year}{2014}), \eprint{1411.2685}.

\bibitem[{\citenamefont{{Ackermann} et~al.}(2013)\citenamefont{{Ackermann},
  {Ajello}, {Albert}, {Allafort}, {Antolini}, {Baldini}, {Ballet},
  {Barbiellini}, {Bastieri}, {Bechtol} et~al.}}]{FAVA}
\bibinfo{author}{\bibfnamefont{M.}~\bibnamefont{{Ackermann}}},
  \bibinfo{author}{\bibfnamefont{M.}~\bibnamefont{{Ajello}}},
  \bibinfo{author}{\bibfnamefont{A.}~\bibnamefont{{Albert}}},
  \bibinfo{author}{\bibfnamefont{A.}~\bibnamefont{{Allafort}}},
  \bibinfo{author}{\bibfnamefont{E.}~\bibnamefont{{Antolini}}},
  \bibinfo{author}{\bibfnamefont{L.}~\bibnamefont{{Baldini}}},
  \bibinfo{author}{\bibfnamefont{J.}~\bibnamefont{{Ballet}}},
  \bibinfo{author}{\bibfnamefont{G.}~\bibnamefont{{Barbiellini}}},
  \bibinfo{author}{\bibfnamefont{D.}~\bibnamefont{{Bastieri}}},
  \bibinfo{author}{\bibfnamefont{K.}~\bibnamefont{{Bechtol}}},
  \bibnamefont{et~al.}, \bibinfo{journal}{\apj} \textbf{\bibinfo{volume}{771}},
  \bibinfo{eid}{57} (\bibinfo{year}{2013}), \eprint{1304.6082}.

\end{thebibliography}

\end{document}